\documentclass[final,5p,times,twocolumn,authoryear, ,table,xcdraw]{elsarticle}

\makeatletter
\def\ps@pprintTitle{%
  \let\@oddhead\@empty
  \let\@evenhead\@empty
  \def\@oddfoot{\reset@font\hfil\thepage\hfil}
  \let\@evenfoot\@oddfoot
}
\makeatother
\usepackage{graphicx}%
\usepackage{multirow}%
\usepackage{amsmath,amssymb,amsfonts}%
\usepackage{amsthm}%
\usepackage{mathrsfs}%
\usepackage[title]{appendix}%
\usepackage{xcolor}%
\usepackage{textcomp}%
\usepackage{manyfoot}%
\usepackage{booktabs}%
\usepackage{algorithm}%
\usepackage{algorithmicx}%
\usepackage{algpseudocode}%
\usepackage{listings}%
\usepackage{tabularx}
\usepackage{datetime}
\usepackage{multirow}
\usepackage{longtable}
\usepackage{lscape}
\usepackage{supertabular,booktabs}
\begin{document}

\begin{frontmatter}

%% Title, authors and addresses

%% use the tnoteref command within \title for footnotes;
%% use the tnotetext command for theassociated footnote;
%% use the fnref command within \author or \affiliation for footnotes;
%% use the fntext command for theassociated footnote;
%% use the corref command within \author for corresponding author footnotes;
%% use the cortext command for theassociated footnote;
%% use the ead command for the email address,
%% and the form \ead[url] for the home page:
%% \title{Title\tnoteref{label1}}
%% \tnotetext[label1]{}
%% \author{Name\corref{cor1}\fnref{label2}}
%% \ead{email address}
%% \ead[url]{home page}
%% \fntext[label2]{}
%% \cortext[cor1]{}
%% \affiliation{organization={},
%%            addressline={}, 
%%            city={},
%%            postcode={}, 
%%            state={},
%%            country={}}
%% \fntext[label3]{}

\title{Gendered Inequalities in Online Harms: Fear, Safety Work, and Online Participation}

\tnotetext[note1]{This is a revised version of a manuscript previously posted to arXiv. The prior version remains available for transparency.}

%%\title{Women are less comfortable expressing opinions online than men and report heightened fears for safety: Measuring gender differences in experiences of online harms}

%% use optional labels to link authors explicitly to addresses:
\affiliation[label1]{organization={Public Policy Programme, The Alan Turing Institute},
city={London},
country={UK}}

\affiliation[label2]{organization={Oxford Internet Institute, University of Oxford},
city={Oxford},
country={UK}}

\affiliation[label3]{organization={Data Science Institute, The London School of Economics and Political Science},
city={London},
country={UK}}

\affiliation[label4]{organization={Department of Sociology, The London School of Economics and Political Science},
city={London},
country={UK}}

%%\affiliation[label2]{organization={OfCom},
%%city={London},
%%country={UK}}

\author[label1]{Florence E. Enock}
\author[label1]{Francesca Stevens}
\author[label1]{Tvesha Sippy}
\author[label1]{Jonathan Bright}
\author[label1]{Miranda Cross}
\author[label1]{Pica Johansson}
\author[label1,label4]{Judy Wajcman}
\author[label1,label2,label3]{Helen Z. Margetts}

\begin{abstract}
%% Text of abstract
Online harms, such as hate speech, trolling and self-harm promotion, continue to be widespread. There are growing concerns that these harms may disproportionately affect women, reflecting and reproducing existing structural inequalities within digital spaces. Using a nationally representative survey of UK adults (N=1992), we examine how gender shapes exposure to a variety of online harms, fears surrounding being targeted, the psychological impact of online experiences, the use of safety tools, and comfort with various forms of online participation. We find that while men and women report roughly similar levels of absolute exposure to harmful content online, women are more often targeted by contact-based harms including image-based abuse, cyberstalking and cyberflashing. Women report heightened fears about being targeted by online harms, more negative psychological impact in response to online experiences, and increased use of safety tools, reflecting more engagement with personal safety work. Importantly, women also say they are significantly less comfortable with several forms of online participation, for example just 23\% of women are comfortable expressing political views online compared to 40\% of men. Explanatory models show direct associations between fears surrounding harms and comfort with particular online behaviours. Our findings show how online harms reinforce gender inequality by placing disproportionate psychological burden and participation constraints on women. These results are important because with much public discourse happening online, we must ensure all members of society feel safe and able to participate in online spaces.
\end{abstract}

%%Graphical abstract
%\begin{graphicalabstract}
%\includegraphics{grabs}
%\end{graphicalabstract}

%%Research highlights
%\begin{highlights}
%\item Research highlight 1
%\item Research highlight 2
%\end{highlights}

\begin{keyword}
%% keywords here, in the form: keyword \sep keyword, up to a maximum of 10 keywords
Online harms \sep Internet safety \sep Social media \sep Gendered online harms \sep Safety work \sep Digital inequality \sep Public attitudes

%% PACS codes here, in the form: \PACS code \sep code

%% MSC codes here, in the form: \MSC code \sep code
%% or \MSC[2008] code \sep code (2000 is the default)
\end{keyword}

\end{frontmatter}

%\tableofcontents

%\linenumbers

%% main text

\section{Background}
\label{background}
\subsection{The problem of online harms} 
While social media has the potential to bring benefits to its users, such as fostering social connection, facilitating education and empowering people with a platform for their voice, social media and the online world more broadly also have the potential to expose people to many forms of harm. Online harms are typically understood as content or activity that is illegal or that risks causing significant physical or psychological harm to individuals, such as hate speech, abuse, misinformation, image-based sexual abuse and self-harm promotion, among many others. Previous work has found exposure to online harms to be high: 66\% of adults in Britain have seen content which they consider to be harmful online, while for younger adults aged 18-34, this is 86\% \citep{Turing2023_harmstracker}. 

\subsection{Gendered experiences of online harms}
Exposure to online harms can be thought of not only as posing a risk to individuals but also as reflecting and reproducing offline structural inequalities. Building on the concept of the `continuum of violence' posed by \citet{Kelly1988}, feminist scholars have more recently argued that digital abuse and harassment are part of the broader spectrum of gendered violence that extends across offline and online contexts \citep[e.g.,][]{HenryPowell2017, jane2017feminist}. There is some empirical evidence to justify concerns that the impact of online harms reflects this digital extension of existing gendered inequalities, with women and girls often disproportionately affected by these experiences. Online survey research carried out by the UK Victims' Commissioner office found that for the majority of categories of online abuse, women were more likely to have been victims than men \citep{Storry&Poppleton2022}. Work has also found that 90\% of those who have been the victim of non-consensual digital distribution of intimate images (sometimes known as `revenge porn') are women \citep{UN2018}. A large-scale review which mapped the state of online violence against women and girls in Europe reported that 9 million girls have experienced some type of online violence by the time they are 15 years old, and across the world, women are 27 times more likely to experience online harassment than men, leading the authors of the report to describe the internet as a place of gendered violence \citep{EWL2017}. This concern is corroborated by recent work finding that more than one in five young men hold a positive view of Andrew Tate, a self-proclaimed misogynist who has posted content advocating for the hitting and choking of women \citep{Booth2024, KCL2024}. 

The evidence described above supports the notion that the online space reflects structural inequalities offline, aligning with feminist accounts of how misogyny adapts across contexts, with online spaces now continuations of gendered social orders rather than neutral arenas \citep[e.g.,][]{banet2016masculinitysofragile, lewis2017online, nadim2019silencing}.  Indeed, further work also suggests that online violence against women often comprises `silencing strategies' such as rape and death threats, attempting to prevent women from partaking in online discussion \citep{lumsden2017media}. Growing concerns over the problem in recent years has led to the development of the UK Government's strategy to tackle violence against women and girls in 2021 \citep{HomeOffice2021}, and the recently passed Online Safety Act now requires Ofcom to develop guidance for tech companies to reduce harm to women and girls \citep{EVAW2023}.

However, there is some conflicting evidence regarding gender differences in absolute exposure to online harms, with some recent work finding no differences in overall exposure \citep{Turing2023_harmstracker, OfCom2023}, and some reports even finding that men experience certain harms such as `sextortion' (online blackmail where criminals threaten to release sexual or indecent images unless money is paid) and physical threats to a greater extent than women \citep{BBC2023, Pew2021}. Explaining this apparent contradiction, one possibility is that much work examining online violence against women in the current literature does not include men in the sample, making it difficult to directly measure how gender interacts with nuanced experiences of online harms and perhaps emphasising the experiences of women \citep[e.g.,][]{Amnesty2017, EWL2017}. Another possibility is that the kinds of harms affecting women online are qualitatively different to those affecting men, with women more at risk of contact-based harms such as sexual harassment and cyberflashing \citep{levrant2001cyberaggression, gillett2023not, EWL2017}, meaning differences may go undetected when asking about exposure to harms as a whole.  In this work, we aim to deepen our understanding of people's key experiences, concerns and behaviours surrounding online harms and online safety with a particular focus on when and how women are disproportionately affected. We examine the impact of gender, exposure to harm and fears surrounding such exposure on online behaviours in a nationally representative sample of adults living in the UK.
 
\subsection{Psychological impacts of online experiences}
Beyond gendered inequalities in exposure to online harms, it is also important to consider their psychological impact. Research has found that exposure to online harms can cause severe harm to the psychological wellbeing of those exposed. One large cross-sectional study of Americans and Finns aged 15-30 found that exposure to harm-advocating content such as eating disorder promotion content, self-harm encouragement and sites portraying deaths of others positively was associated with significantly lower subjective wellbeing \citep{keipi2017harm}. Further, a systematic review found that victims of cyberstalking and harassment commonly reported experiencing depression, fear, anger, suicidal ideation, shame, paranoia and isolation, amongst other negative emotions \citep{stevens2021cyber}. As well as potentially facing greater targeting of certain online harms, women may also suffer more psychologically as a result of such exposure because of the way these abuses operate as forms of `symbolic violence' (described as the normalised ways in which inequality is reproduced and accepted \cite{bourdieu1991language}). Through this lens, gendered abuse online is not only damaging to individuals but also reinforces women’s unequal position within a patriarchal society. Supporting this, research by Amnesty found that of the 20\% of women who had experienced abuse or harassment through social media, 55\% suffered stress, anxiety or panic attacks and 61\% had trouble sleeping \citep{Amnesty2017}. Additionally, only 3\% of women who had experienced abuse or harassment say they were untroubled by these experiences (compared to 17\% of men), while 49\% of women (compared to 35\% of men) stated that their experiences of online abuse made them feel ashamed \citep{Storry&Poppleton2022}. Recent work from Ofcom showed that women were less likely than men to say that being online has a positive effect on their mental health (40\% of women compared to 45\% of men) \citep{OfCom2022}. While some prior work in journalism and media studies has examined the negative psychological impact of online abuse on women \citep[e.g.][]{chen2020you}, few studies have systematically compared the psychological experiences of men and women in response to specific online experiences in a representative sample. To address this gap, we examine whether women are more likely than men to be negatively emotionally affected by online experiences.

\subsection{Fears surrounding online experiences}
Closely tied to negative psychological experience is heightened fear. Offline, women consistently report higher levels of fear of being targeted by crime compared with men, especially when considering crime relating to personal harm \citep{snedker2012explaining}. \cite{stanko1995women} argues that women’s concerns around danger are largely caused by a fear of men, reflecting women's positions in a gendered world. It is important to understand feelings of fear around safety online as these are likely to impact people's comfort with participating in the online space. While some work suggests that women's fears around personal safety are heightened compared to men's, few previous studies have systematically examined gender differences in fears of specific types of online harm, which is a key aim of the present study. We also explore where fears of online harms typically originate from, an important first step in understanding the function of cultural transmission in discussions about online safety.

\subsection{Online safety work and participation}
Fears surrounding online harms can be understood as part of a broader pattern of gendered safety work, where the burden falls upon women to manage their own risks \citep{VeraGrayKelly2020}.  In an online context, such safety work often requires engaging with platform safety technologies, such as blocking, reporting, or altering privacy settings \citep{bright2024understanding}. However, little research has explored whether there are gender differences in uptake of such tools, which would align with theoretical claims that women engage in disproportionate self-protection work online. We examine whether women are more likely than men to engage in safety work online, reflected in increased use of safety tools. 

Offline, fear of crime regularly leads to changes in behaviour. As fear of crime is often higher in women, the restrictive behavioural consequences can be greater for women too \citep{stanko1990everyday, snedker2012explaining}. This can impact women in a myriad of ways, for example in the extra measures they take to ensure they get home safely at night and additional precautions when travelling alone \citep{ONS2022}. Just as women work harder to protect their safety offline, it looks increasingly likely that this additional burden must be carried into the online world too. Work from Ofcom (2022) found that women are significantly less likely than men to feel that they are able to have a voice online, while research exploring academics' interactions with social media found that fears of online harassment in female and ethnic minority academics often leads to self-censorship \citep{olson2018combating}. These findings show the silencing effect that online abuse can have, with such harms directly threatening freedom of expression. In the digital context, this extra work can place uneven demands on women and sustain digital inequalities \citep{vera2018right}. We investigate the impact of gender on comfort with online participation, such as expressing political opinions and sharing content online, and test whether fears of online harms are associated with reduced comfort with online participation. 

%I am removing the below for now because it's less relevant now we don't include exposure as a predictor, and this section is very long already. Saving it in case we change our minds following a review. 
%Changes in behaviour may also occur as a direct result of experiencing harm. Indeed, prior work finds that women conduct themselves differently online after having experienced abuse, with over three quarters of women who had experienced abuse or harassment on a social media platform subsequently altering how they used social media, including a third revealing that they no longer posted content expressing opinions \citep{Amnesty2017}. Plan International found that one in five girls who have been harassed online either stopped using the platform after the abuse, or considerably decreased the amount of time spent on it \citep{Plan2020}. Work elsewhere supports accounts that victims of online abuse may withdraw from technology and social media as a result of their experiences, isolating them further \citep{watson2023investigating, stevens2021cyber}. 

\section{Research contribution}
We seek to address the gaps in the research landscape that remain surrounding when and how women are disproportionately affected by harmful content online. We build on feminist accounts that frame online harms as part of broader patterns of gendered inequality, and ask when and how such harms affect women differently to men.

While much existing work examining online violence against women and girls includes only women and girls in the sample as the key subjects of interest, this means it can be difficult to understand the true extent to which women are affected differently to men. We survey a sample of UK adults nationally representative across age, gender and ethnicity so that we can make meaningful gender comparisons across our variables of interest. Additionally, while much of the existing research either asks about exposure to harmful content in general, or focuses on very specific harms such as abuse and harassment, we ask about exposure and fears relating to a comprehensive list of fifteen potential harms, including content-based ones (those that arise from viewing certain pieces of content) like misinformation and eating disorder promotion content, along with contact-based ones (those that originate from direct behaviours towards the target) such as cyberflashing, cyberstalking and image-based sexual abuse. Importantly, we ask not only about general exposure to such harms, but also about whether people have been directly targeted by each one and to what extent. In doing so, we are able to capture a more detailed picture of the different ways in which men and women experience different types of harmful content online. 

Our survey design also allows us to deepen our understanding of when and how women are affected differently by online harms to men. Much previous work in the area looks only at how feelings and behaviours change after harm exposure, and only in samples that report having been affected. In our sample, we are able to compare psychological impacts of online experiences between men and women directly, along with fears, the origins of such fears, and subsequent behavioural changes. We are also able to understand whether women engage in more work overall to protect their safety online, including censoring themselves or behaving more cautiously. We also examine associations between gender, fears and online participation.

\section{Methodology}
\label{Data and Methods}
\subsection{Data collection and ethics} 
We conducted a nationally representative survey of 2,000 UK adults. Data collection took place online during June 2023, the survey was created and administered using Qualtrics\footnote{www.qualtrics.com} and participants were recruited through Prolific\footnote{www.prolific.com}.The survey was approved by the Ethics Committee at The Alan Turing Institute, UK (approval number C2105-074). Informed consent was obtained at the start of the survey according to approved ethical procedures. The materials and data are available on request.

\subsection{Sample} A total of 1992 participants who completed the survey passed standard checks for data quality and were included in the final sample. The sample was designed to be nationally representative of the population of the United Kingdom across demographic variables of age, gender and ethnicity using Prolific’s representative sample tool. Respondents were between 18 and 90 years old, with a mean age of 45.8 (SD = 15.5). A total of 1010 participants identified as female (50.7\%), 965 as male (48.4\%), with 9 as non-binary, 5 selecting ‘prefer not to say’, and 3 self-describing. The majority of respondents identified as White (86.9\%), while 7.2\% as Asian or Asian British, 3.0\% as Black, African, Caribbean, or Black British, and 1.6\% as mixed, multiple or other ethnicities (0.7\% participants selected ‘any other ethnic group’, while 0.6\% chose ‘prefer not to say’).\footnote{Although participants indicated more specific ethnic identities, we have combined them into broader categories to simplify reporting here.} As the main aim of our survey is understanding gender differences, we exclude the 17 participants choosing gender as 'non-binary', 'prefer not to say' or 'prefer to self-describe' from these analyses because the sample size is too small to make meaningful comparisons and inferences for these sub-samples. 

\subsection{Survey} 
\textbf{Demographics and background questions}: For each participant, we collected standard demographic information about age, gender, ethnicity, education and political orientation. Age could be entered as any number with a minimum of 18. For gender, ethnicity and education level, participants were asked to select the option that they felt best described them from a list of standard predefined categories. For political orientation, participants were asked to select the option that they felt most described their beliefs: `More to the left', `Centre', `More to the right' and `Prefer not to say'. Participants were also asked what kind of device they were using to complete the survey, as well as how often they use the internet, whether they use/have used social media before, and how often. All demographic questions, other than age, provided participants with a ‘prefer not to say’ option.

\textbf{Exposure to online harms}: We presented participants with a list of 15 online harms along with definitions for each. These harms were selected to be as comprehensive as possible and to reflect categories of interest in UK policy, drawing on lists developed by Ofcom and the Victims Commissioner \citep{OfCom2022, Storry&Poppleton2022}. The harms we included were: Bullying, Catfishing, Cyberflashing, Cyberstalking, Doxing, Eating disorder promotion content, Group attacks, Hate speech, Image-based sexual abuse, Impersonation, Misogyny, Misinformation, Threats of non-sexual violence, Threats of sexual violence, and Trolling\footnote{We note here that while the harms we included are defined separately and analysed as distinct categories, in practice they may overlap (for example, misinformation can contain hate, bullying can contain misogyny and so on). It is possible, then, that participants could have selected multiple harms to describe a single experience. Our analyses do not require the harms to be conceptually independent and the overlap across categories reflects the interconnected nature of online harms.} (see Table \ref{tab:definitions} in Supplementary Information for the definitions we provided to participants). Following prior work \citep{Turing2023_harmstracker}, participants were asked to indicate whether they had heard of or seen each harm online in the past year (Never heard of/ Heard of but not seen in the last year/ Seen online in the last year). 

For each harm, if participants indicated they had seen it in the past year, they were then asked a follow up about the extent to which they had directly received each harm and the extent to which they had witnessed each harm. For `content-based' harms (hate speech, misinformation, misogyny and eating disorder promotion content), `directly received' meant the content was directly intended for the participant, for example naming them or being sent in a direct message, while for `contact-based' harms (trolling, bullying, stalking, cyberflashing, group attacks, catfishing, threats of physical violence, threats of sexual violence, doxing and image-based sexual abuse), `directly received' meant the harm happened to them. In the case of impersonation, directly received meant that someone impersonated the participant's account. `Witnessing' always meant that they had seen the content but that it was not directly intended for them personally. Response options were: Never/ Once/ 2-4 times/ 5 or more times/ Prefer not to say. 

For each harm, if participants indicated they had directly received it, they were then asked whether they knew the identity of the person/people responsible (Response options: Yes, it was someone/people I know/am familiar with (either from prior offline or online encounters)/ Yes, I was able to identify them but they were previously unknown to me / No, it was from an anonymous account(s) or with a fake name / A mixture of any of the above / Other, please specify / I cannot remember).

\textbf{Fears of exposure to online harms}: To measure fears about exposure to each of the same 15 harms, participants were asked to indicate the extent to which they fear witnessing such content, and the extent to which they fear directly receiving such content (both scales: Not at all / Not very much / Somewhat / Very much). This measure was adapted from standard approaches in the fear-of-crime literature, which often assess intensity of fear using similar Likert scales \citep[e.g.,][]{callanan2009exploration}. We note here that we did not define fear for participants and so, consistent with the fear-of-crime literature, responses could reflect emotional reactions to the possibility of harm exposure, more cognitive risk assessments of such exposure, or a combination of the two. 

If participants indicated any level of fear about receiving any of the harms, they were asked where they thought their these fears might have come from (could choose as many as apply from: Personal past experience/ Female friend's experience/ Male friend's experience/ Non-binary friend's experience/ From the media/ Public figure's experience/ Something else/ Not sure). Participants were also asked how, if at all, fears about receiving online harms affect their online behaviours (could choose as many as apply from: Less likely to use social media platforms in general/ Less likely to share opinions online/ Less likely to share photographs of themselves online/ Less likely to share content online in general/ Does not affect behaviour online/ Something else (please specify)/ Not sure).

\textbf{Comfort with online participation}: Participants were asked how comfortable they are (Not at all/ Not very/ Somewhat/ Extremely) engaging with seven different online activities in both public/open settings and private/restricted settings. The behaviours we asked about were: Expressing political opinions online; Expressing other opinions online (e.g., opinions about the news, TV shows or music); Challenging content they see online and do not agree with; Sharing personal information online (such as name, date of birth, where they live and work, gender, religious beliefs, and so on); Sharing photos of themselves online; Sharing photos of friends and family online; Sharing photos of activities online (e.g., travelling, cooking or going out with friends). Items for the comfort measures were developed specifically for this study but were informed by prior work on online voice and participation \cite{OfCom2022, olson2018combating, papacharissi2015affective}. 

\textbf{Psychological impact}: Participants were asked to indicate the extent to which certain experiences online had caused them particular kinds of feelings. These six feelings were: Feeling sad or low; Feeling angry or frustrated; With physical symptoms (such as insomnia, headaches and stomach aches); Feeling as though their job/career had been negatively affected; Feeling as though relationships with a partner, friends or family had been negatively affected; Feeling (more) eager or keen to use social media to advocate for a specific cause, political ideology or to educate others on a specific topic (Response scales for all were: Not at all/ Not very much/ Somewhat/ Very much). We drew on prior research to inform the items we included \citep{Amnesty2017, stevens2021cyber}, but the set of six items used was developed specifically for this study. 

\textbf{Engagement with safety tools}: Drawing on recent survey work examining uptake of online safety technology \citep{bright2024understanding}, participants were asked to indicate if they had ever used seven types of online safety features on social media platforms. These were: Disabled location sharing on a device; Disabled airdrop and/or Bluetooth on a device; Made profile/account/page private; Limited who can contact you (for example so you cannot receive messages from people you don't know); Limited who can engage with your contents (for example who can like or comment on your posts); Limited who can tag/mention you in images/posts/tweets; Limited how people can find your profile (for example controlling whether people can find your profile by linking with email or phone, through a search engine or by appearing in suggested people lists) (Yes/No for all). 

\subsection{Procedure}
After participants gave their informed consent to take part in the survey, they responded to the demographic questions, questions about device use, time spent online, and social media use. Following this, they read brief definitions of the fifteen harms so as to familiarise themselves with what each one means. They were asked to pay close attention to the definitions and could move on after a minimum of 45 seconds. Participants then responded to the questions about harm exposure for each of the fifteen harms (awareness/exposure, followed by extent of exposure and perpetrator identity if relevant), followed by questions about fears (including fear origins and subsequent behavioural changes if relevant). Next, participants answered questions about comfort with various forms of online participation, the psychological impact of online experiences, and engagement with safety tools online. At the end of the questions, participants were given an opportunity to provide feedback in a free text box before continuing to the debrief and finally completing the submission and being returned to Prolific for payment. The survey was designed to take approximately 15 minutes to complete and each participant received £2.25 for their time.

\subsection{Data analysis} 
We present descriptive statistics for overall proportions of men and women choosing each response option for each question. We use logistic regressions and non-parametric between samples t-tests (Mann-Whitney U-Tests) to test for gender differences in each of the outcome variables (information about specific outcome variables and tests are in the relevant part of Results). 

Additionally, we test for associations between fears about online harms and comfort engaging in several forms of online participation by running a series logistic regression analyses, each including fear as predictor (one for fear about each of 12 harms, see Results for harms included), and comfort engaging in the online behaviour of interest as the outcome (one for comfort with each of the seven behaviours), all controlling for gender, age, and internet use. For all statistical analyses, the accepted significance threshold was set at .05. 

\section{Results}
\label{Results}
\subsection{Experiences of online harms}
\subsubsection{Awareness and exposure}
We first present responses for awareness of and exposure to the 15 online harms we asked about for men and women. Here, we count exposure by a response indicating `Seen online in the last year'. Logistic regression analyses tested differences in exposure between men and women for each of the 15 harms. 

Gender differences in exposure to various harms show mixed results. Women are 22\% more likely to say they have seen online misogyny (p=.027), 24\% more likely to say they have seen bullying (p=.017) and 104\% more likely to say they have seen eating disorder promotion content (p$<$.001) than men. However, women are 20\% less likely to say they have seen hate speech (p=.016), 25\% less likely to say they have seen misinformation (p=.017), 23\% less likely to say they have seen seen impersonation (p=.007), 31\% less likely to say they have seen threats of physical violence (p$<$.001), and 29\% less likely to say they have seen doxing (p=.003) than men. For many of the harms, such as trolling and cyberstalking, differences in self-reported exposure between men and women are not statistically significant.

Table \ref{tab:Exposure_Harms} shows proportions of men and women indicating they have seen each type of harm in the past year, with asterisks marking significant differences. 

% Please add the following required packages to your document preamble:
% \usepackage{multirow}
\begin{table}
\begin{tabular}{llll}
\textbf{Harm} & \textbf{Gender} & \textbf{\%} & \textbf{Sig.} \\ \hline
Hate Speech & Female & 56.9\% & * \\
 & Male & 62.3\% &  \\ \hline
Misinformation & Female & 81.6\% & * \\
 & Male & 85.6\% &  \\ \hline
Misogyny & Female & 53.0\% & * \\
 & Male & 48.0\% &  \\ \hline
Trolling & Female & 62.3\% &  \\
 & Male & 63.4\% &  \\ \hline
Bullying & Female & 42.9\% & * \\
 & Male & 37.6\% &  \\ \hline
Cyberstalking & Female & 11.1\% &  \\
 & Male & 11.7\% &  \\ \hline
Cyberflashing & Female & 8.7\% &  \\
 & Male & 8.5\% &  \\ \hline
Group attacks & Female & 19.6\% &  \\
 & Male & 22.2\% &  \\ \hline
Impersonation & Female & 31.4\% & ** \\
 & Male & 37.2\% &  \\ \hline
Catfishing & Female & 22.7\% &  \\
 & Male & 25.3\% &  \\ \hline
Physical threats & Female & 26.8\% & *** \\
 & Male & 34.8\% &  \\ \hline
Sexual threats & Female & 16.5\% &  \\
 & Male & 18.3\% &  \\ \hline
Doxing & Female & 15.9\% & ** \\
 & Male & 21.1\% &  \\ \hline
\multirow{2}{*}{\begin{tabular}[c]{@{}l@{}}Image-based \\ sexual abuse\end{tabular}} & Female & 8.4\% &  \\
 & Male & 8.5\% &  \\ \hline
\multirow{2}{*}{\begin{tabular}[c]{@{}l@{}}Eating disorder \\ content\end{tabular}} & Female & 27.1\% & *** \\
 & Male & 15.4\% & 
\end{tabular}
 \caption{Proportions of men and women indicating they have seen each harm in the past year. In the Sig. column, * means significant at the p$<$.05 level; ** at the p$<$.01 level, and *** at the p$<$.001 level.}
    \label{tab:Exposure_Harms}
\end{table}

\subsubsection{Extent of exposure}
Importantly, the results on exposure presented above show whether or not people have seen certain types of content but not whether they have been direct targets of online harms, nor how much. To understand the extent of people's exposure more deeply, for each of the 15 harms that participants indicated they had seen in the past year, we analysed responses to follow up questions about extent to which they had witnessed such content and the extent to which they had directly received such content (see Methods). Response options for both scales were converted into 4 point numeric scales (1 = Never; 2 = Once; 3 = 2-4 times; 4 = 5 or more times, with `Prefer not to say' as NA) and non-parametric t-tests (Mann-Whitney U-Tests) examined gender differences in being directly targeted by each type of harm.\footnote{An alternative analysis is to create binary responses (Never = 0 / At least once = 1) and use logistic regressions to examine the effect of gender on whether people have been targeted before or not. Taking this approach yields identical results.} If participants had previously indicated that they had not seen the harm in the past year and were therefore not asked follow-up questions about the capacity in which they had seen it, their responses for this question were re-coded from missing to `Never'. 

\begin{figure*}[ht]
    \centering
    \includegraphics[width=1\linewidth]{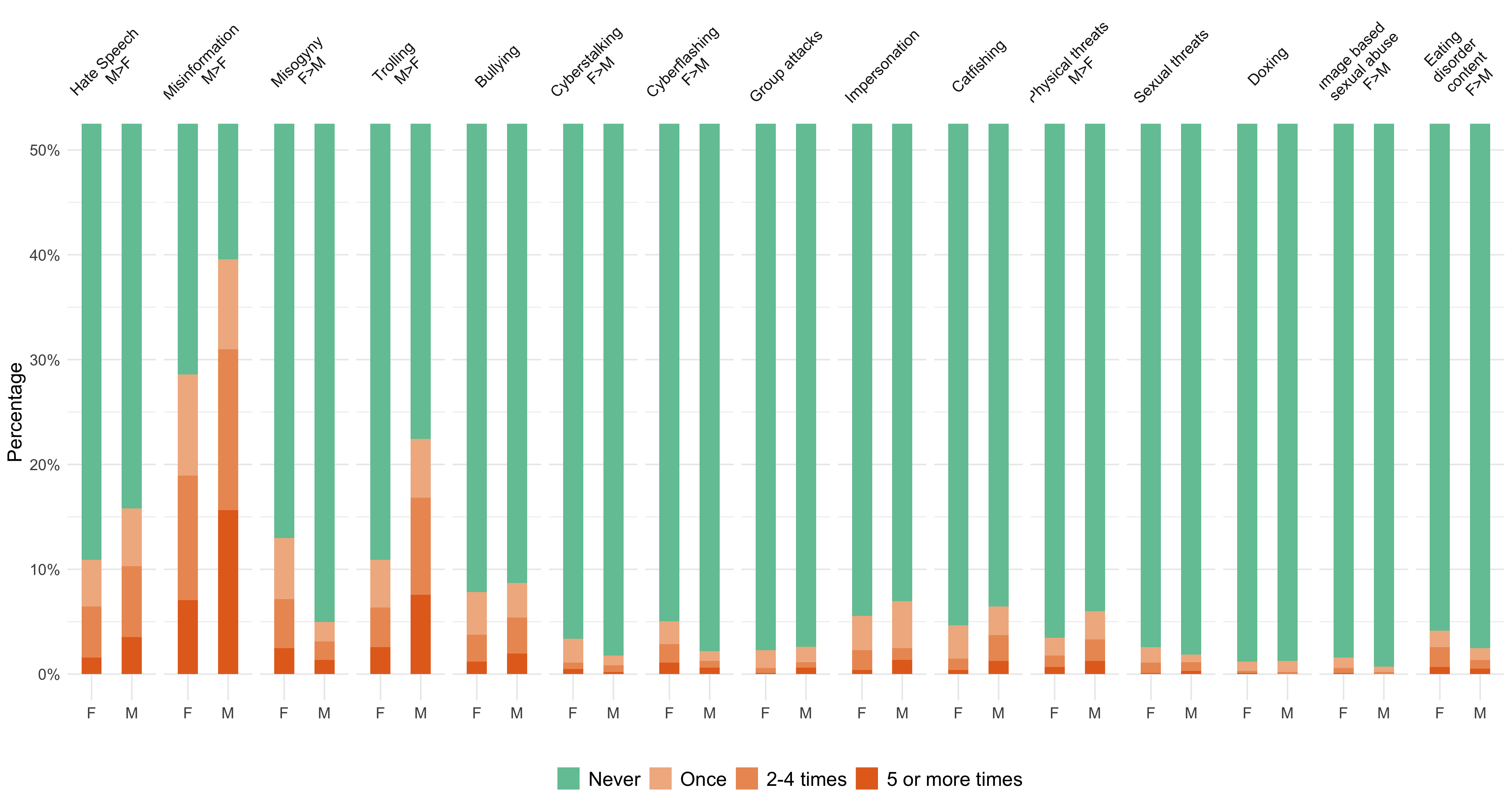}
    \caption{The extent to which people report having been direct targets of each online harm by gender. F$>$M indicates women are targeted significantly more than men. M$<$F indicates men are targeted significantly more than women. `Prefer not to say' responses were uncommon and are not included.}
    \label{fig:harm_targeted}
\end{figure*}

Women reported being the direct targets of several online harms to a significantly greater extent than men, including online misogyny (p$<$.001), cyberstalking (p=.026), cyberflashing (p$<$.001), eating disorder promotion content (p=.039) and, marginally, image based abuse (p=.075). Table \ref{tab:Exposure_DR_women} in Supplementary Information shows proportions of men and women choosing each response option for the five harms that women report being targeted by to a greater extent than men. Men reported being the direct targets of hate speech (p$<$.001), misinformation (p$<$.001), trolling (p$<$.001) and threats of physical violence (p=.007) to a greater extent than women. Table \ref{tab:Exposure_DR_men} in Supplementary Information shows proportions of men and women choosing each response option for the four harms that men report being targeted by to a greater extent than women. There were no significant gender differences in reports of being directly targeted by the remaining six harms. Figure \ref{fig:harm_targeted} shows the extent to which men and women report having been direct targets of each online harm. Note that while we asked individuals who had been targeted by harms whether they knew the identity(ies) of the perpetrator(s), comparing proportions of responses led to sample sizes too small to make meaningful comparisons for this question.

% NOTES: The results below were only on the subsample of people who had seen the harm. So, of the people who had seen image based abuse, women were more likely to be direct targets than men. This is slightly different from new version which is overall, are women more likely to be direct targets than men. Results change a little bit: 
%Women reported being the direct targets of several online harms to a significantly greater extent than men, including online misogyny (p$<$.001), cyberstalking (p=.010), cyberflashing (p$<$.001), threats of sexual violence (p=.052) and image based abuse (p=.011). Table \ref{tab:Exposure_DR_women} shows proportions of men and women choosing each response option for the five harms that women report being targeted by to a greater extent than men. Men reported being the direct targets of hate speech (p=.006), misinformation (p$<$.001) and trolling (p$<$.001) to a greater extent than women. Table \ref{tab:Exposure_DR_men} shows proportions of men and women choosing each response option for the three harms that men report being targeted by to a greater extent than women. There were no significant gender differences in reports of being directly targeted by bullying, group attacks, impersonation, catfishing, threats of (non-sexual) physical violence, doxing, and eating disorder content.

Comparing gender differences in witnessing each type of harm (an individual seeing harmful content online not directly targeted at them), men reported witnessing hate speech, misinformation, trolling, impersonation, catfishing, threats of physical violence, threats of sexual violence, doxing, and image-based abuse to a greater extent than women (all ps$<$.05), while women reported witnessing eating disorder promotion content (p$<$.001) and, marginally, misogyny (p=.051) and bullying (p=.073) to a greater extent than men.\footnote{We explored whether possible gender differences in overall internet use explained these results but we found men and women report using the internet and social media to a similar extent overall.} The differences here in results for witnessing and directly receiving online harms shows the importance of understanding both who is most likely to be targeted, alongside understanding overall exposure. 

\subsection{Psychological impact of online experiences}
We asked participants the extent to which experiences online had ever caused them six particular kinds of feelings (see Methods). For each of these responses, we created a binary outcome variable to indicate whether they had experienced this psychological effect or not (Yes, 1 - Somewhat/Very much and No, 0 - Not at all/Not very much). Overall, 48\% of respondents said they had been left feeling sad or low as a result of an online experience; 68\% that they had felt angry or frustrated; 11\% with physical symptoms (such as insomnia, headaches and stomach aches); 9\% as though their career had been negatively affected; 15\% as though their relationships had been negatively affected; and 22\% feeling (more) eager or keen to use social media to advocate for a specific cause. Logistic regression analyses with gender as predictor and response as outcome for each of the items found that women were significantly more likely than men to have been left feeling sad or low (96\% more likely), angry or frustrated (54\% more likely), and with physical symptoms (47\% more likely) as a result of an experience online (all ps$<$.001). There were no gender differences in responses for the remaining three questions about psychological impact (all ps $>$.05). Table \ref{tab:Feelings} in Supplementary Information shows responses to each item for men and women. 

\subsection{Fears about exposure to online harms}
\subsubsection{Overall extent of fears about receiving online harms}
To compare gender differences in fears about exposure to each of the 15 online harms, we created a binary outcome for fear, with `Somewhat' and `Very much' as 1 - Fearful, and `Not at all' and `Not very much' as 0 - Not fearful. Women consistently express significantly greater levels of fear than men across all 15 types of content, both for witnessing and for directly receiving each harm. Women are 117\% more likely to fear receiving hate speech than men, 39\% more likely to fear receiving misinformation, 489\% more likely to fear receiving misogyny, 125\% more likely to fear being targeted by trolling and bullying, 120\% more likely to fear being targeted by cyberstalking, 196\% more likely to fear being targeted by cyberflashing, 68\% more likely to fear being targeted by group attacks, 50\% more likely to fear having an account impersonated, 64\% more likely to fear catfishing, 112\% more likely to fear receiving physically violent threats, 215\% more likely to fear receiving sexually violent threats, 38\% more likely to fear being targets of doxing, 133\% more likely to fear being targets of image-based abuse, and 174\% more likely to fear receiving eating disorder promotion content (all ps$<$.001). Table \ref{tab:Fears} shows proportions of men and women fearing receiving each type of online harm. 
\begin{table}
\begin{tabular}{llll}
\textbf{Harm} & \textbf{Response} & \textbf{Male} & \textbf{Female} \\ \hline
Hate speech & Low fear & 74.9\% & 57.9\% \\
 & High fear & 25.1\% & 42.1\% \\ \hline
Misinformation & Low fear & 63.1\% & 55.2\% \\
 & High fear & 36.9\% & 44.8\% \\ \hline
Misogyny & Low fear & 86.2\% & 51.5\% \\
 & High fear & 13.8\% & 48.5\% \\ \hline
Trolling & Low fear & 71.5\% & 52.8\% \\
 & High fear & 28.5\% & 47.2\% \\ \hline
Bullying & Low fear & 68.7\% & 49.4\% \\
 & High fear & 31.3\% & 50.6\% \\ \hline
Cyberstalking & Low fear & 66.0\% & 46.9\% \\
 & High fear & 34.0\% & 53.1\% \\ \hline
Cyberflashing & Low fear & 79.0\% & 55.9\% \\
 & High fear & 21.0\% & 44.1\% \\ \hline
Group attacks & Low fear & 69.1\% & 57.1\% \\
 & High fear & 30.9\% & 42.9\% \\ \hline
Impersonation & Low fear & 59.1\% & 49.0\% \\
 & High fear & 40.9\% & 51.0\% \\ \hline
Catfishing & Low fear & 72.2\% & 61.4\% \\
 & High fear & 27.8\% & 38.6\% \\ \hline
Physical threats & Low fear & 67.7\% & 49.7\% \\
 & High fear & 32.3\% & 50.3\% \\ \hline
Sexual threats & Low fear & 74.0\% & 47.4\% \\
 & High fear & 26.0\% & 52.6\% \\ \hline
Doxing & Low fear & 60.7\% & 52.9\% \\
 & High fear & 39.3\% & 47.1\% \\ \hline
\multirow{2}{*}{\begin{tabular}[c]{@{}l@{}}Image based \\ sexual abuse\end{tabular}} & Low fear & 71.0\% & 51.2\% \\
 & High fear & 29.0\% & 48.8\% \\ \hline
\multirow{2}{*}{\begin{tabular}[c]{@{}l@{}}Eating disorder \\ content\end{tabular}} & Low fear & 87.3\% & 71.4\% \\
 & High fear & 12.7\% & 28.6\%
\end{tabular}
 \caption{Level of fear reported by men and women about directly receiving each online harm. Women are significantly more fearful than men about receiving all fifteen types of harm.}
    \label{tab:Fears}
\end{table}

% The below is how we described results before, choosing a few key stats (these would need to be updated now though) - this is another option instead of the table: 
%For example, 49\% of women fear receiving misogynistic content compared to 14\% of men, 54\% of women fear being the target of cyberstalking compared to 35\% of men, 45\% of women fear being targeted by cyberflashing compared to 21\% of men, and 50\% of women fear being the target of image based sexual abuse compared to 30\% of men. These results translate as women being 505\% more fearful of receiving misogynistic content, 119\% more fearful of being the target of cyberstalking, 197\% more fearful of being targeted by cyberflashing, and 136\% more fearful of being targeted by image based sexual abuse than men. Figure 2 shows self-reported fear about receiving each of the 15 harms by gender, with all gender differences statistically significant at the .05 level. 

\begin{figure*}[ht]
    \centering
    \includegraphics[width=1\linewidth]{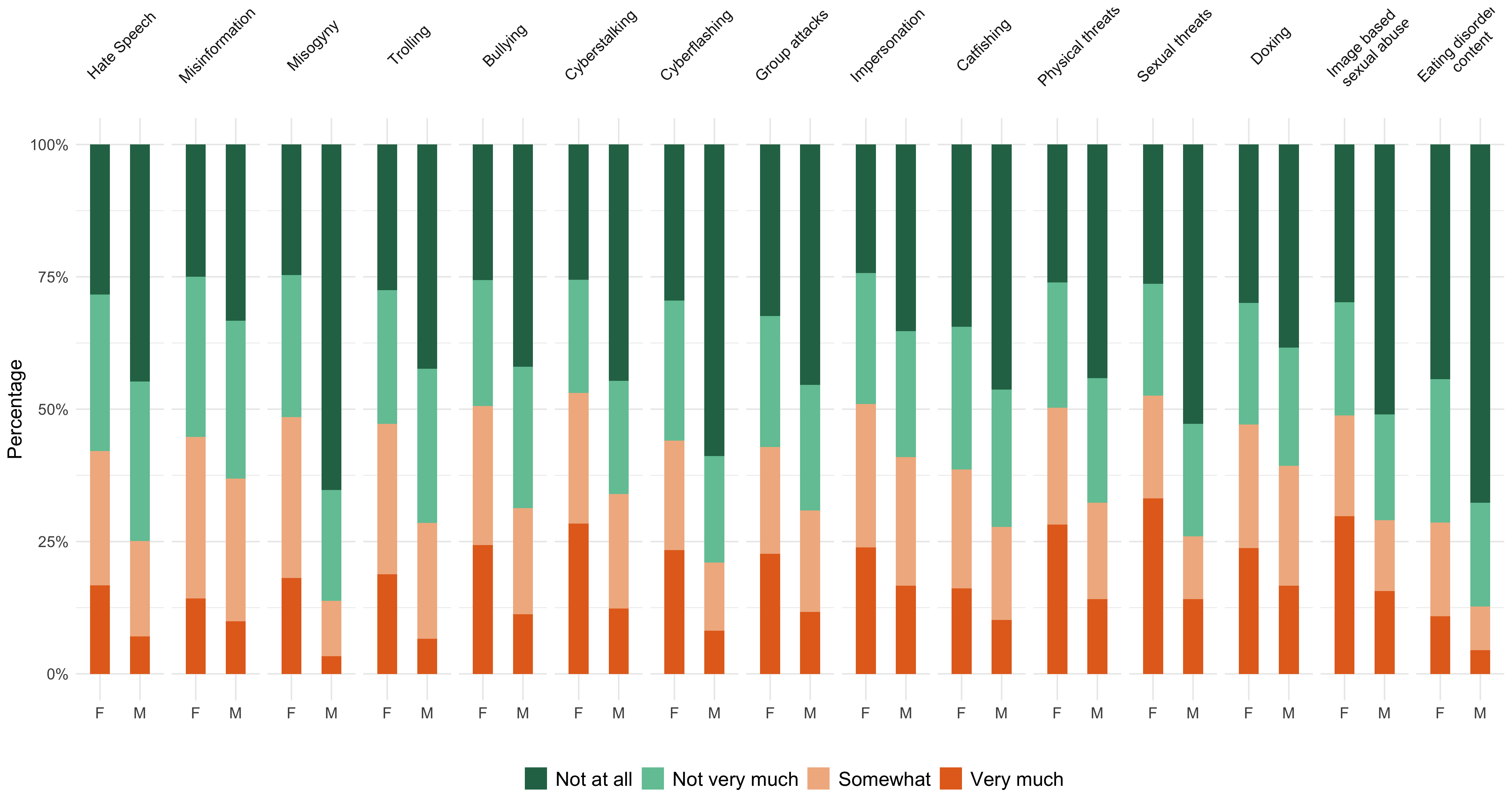}
    \caption{Self-reported fear of receiving harmful content online by gender. All gender differences are statistically significant at the .05 level.}
    \label{fig:fear-recieving}
\end{figure*}

\subsubsection{Origins of fears about receiving online harms}
When asked where they thought fears about receiving online harms might have come from, the top three choices for women were: The media (65\%), Female friend's experience (37\%) and Public figure (35\%). Additionally, 30\% of women chose Personal experience, 8\% chose Male friend's experience, 5\% Non-binary friend's experience. The top three choices for men were: The media (60\%), Personal experience (29\%), and Female friend's experience (27\%). Additionally, 26\% of men chose Public figure, 14\% chose Male friend's experience, and 3\% Non-binary friend's experience. Therefore, both men and women chose the media and female friends' experiences as the most common sources of fears of online harms.

\subsubsection{Behavioural effects of fears}
When asked how, if at all, fears about receiving online harms affect online behaviours, the top three choices for women were: Less likely to share opinions (59\%), Less likely to share photos (54\%) and Less likely to share content in general (46\%). Just 9\% said fears had no effect on online behaviours. The top three choices for men were: Less likely to share opinions (41\%), Less likely to share photos (40\%) and Less likely to share content in general (36\%). 13\% said fears had no effect on online behaviours. 

While both men and women most typically indicated they were less likely to share opinions, photos and general content as a result of fearing being targeted by online harms, these behavioural effects were significantly more pronounced for women. Logistic regression analyses showed that women were 104\% more likely to say they shared opinions less, 78\% more likely to say they shared photos less, and 51\% more likely to say they shared content less in general than men (all ps $<$.001). There was no difference between men and women in whether they were less likely to use social media overall, p=.107). Additionally, women were 30\% less likely than men to say their fears do not affect their behaviour online (p=.015).

\subsection{Safety tools}
 We used logistic regression analyses to compare the use of each of the seven safety tools (Yes=1, No=0) between men and women. Women consistently expressed significantly higher use of all safety tools included. Women were 29\% more likely to have disabled location sharing than men (p=.034); 26\% more likely to have disabled airdrop/bluetooth than men (p=.019); 94\% more likely to have made an account private (p$<$.001); 55\% more likely to have limited who can contact then (p$<$.001); 50\% more likely to have limited who can engage with their content (p$<$.001); 60\% more likely to have limited who can tag them in posts (p$<$.001); and 34\% more likely to have limited how people can find them online (p=.002). See table \ref{tab:Safety_tool_use} for proportions of women and men using each of the seven safety tools.

\begin{table}
\begin{tabular}{llll}
\textbf{Safety tool} & \textbf{Gender} & \textbf{Response} & \textbf{\%} 
\\ \hline
\multirow{4}{*}{Disabled location sharing}  
& Female          & Yes               & 84.5\%      \\
&                 & No                & 15.5\%      \\
& Male            & Yes               & 80.8\%      \\
&                 & No                & 19.2\%      \\ \hline
\multirow{4}{*}{\begin{tabular}[c]{@{}l@{}}Disabled airdrop \\ and/or Bluetooth\end{tabular}}        
& Female          & Yes               & 72.8\%      \\
&                 & No                & 27.2\%      \\
& Male            & Yes               & 67.9\%      \\
&                 & No                & 32.1\%      \\ \hline
\multirow{4}{*}{\begin{tabular}[c]{@{}l@{}}Made profile/account/\\ page private\end{tabular}}        
& Female          & Yes               & 88.9\%      \\
&                 & No                & 11.1\%      \\
& Male            & Yes               & 80.5\%      \\
&                 & No                & 19.5\%      \\ \hline
\multirow{4}{*}{\begin{tabular}[c]{@{}l@{}}Limited who can\\ contact you\end{tabular}}               
& Female          & Yes               & 84.2\%      \\
&                 & No                & 15.8\%      \\
& Male            & Yes               & 77.6\%      \\
&                 & No                & 22.4\%      \\ \hline
\multirow{4}{*}{\begin{tabular}[c]{@{}l@{}}Limited who can \\ engage with your content\end{tabular}} 
& Female          & Yes               & 76.7\%      \\
&                 & No                & 23.3\%      \\
& Male            & Yes               & 68.7\%      \\
&                 & No                & 31.3\%      \\ \hline
\multirow{4}{*}{\begin{tabular}[c]{@{}l@{}}Limited who can \\ tag/mention you in posts\end{tabular}} 
& Female          & Yes               & 70.3\%      \\
&                 & No                & 29.7\%      \\
& Male            & Yes               & 59.7\%      \\
&                 & No                & 40.3\%      \\ \hline
\multirow{4}{*}{\begin{tabular}[c]{@{}l@{}}Limited how people can \\ find your profile\end{tabular}} 
& Female          & Yes               & 70.5\%      \\
&                 & No                & 29.5\%      \\
& Male            & Yes               & 64.1\%      \\
&                 & No                & 35.9\%     
\end{tabular}
 \caption{Proportions of men and women that reported using each of the seven safety tools. Women were significantly more likely to report using every tool.}
    \label{tab:Safety_tool_use}
\end{table}
 
\subsection{Comfort with online behaviours}
To test for gender differences in comfort with seven common online behaviours (listed in Methods), we created a binary outcome for comfort with each behaviour, with `Somewhat' and `Extremely' as 1, Comfortable, and `Not at all' and `Not very' as 0, Not comfortable. 

\begin{figure*}[ht]
    \centering
    \includegraphics[width=1\linewidth]{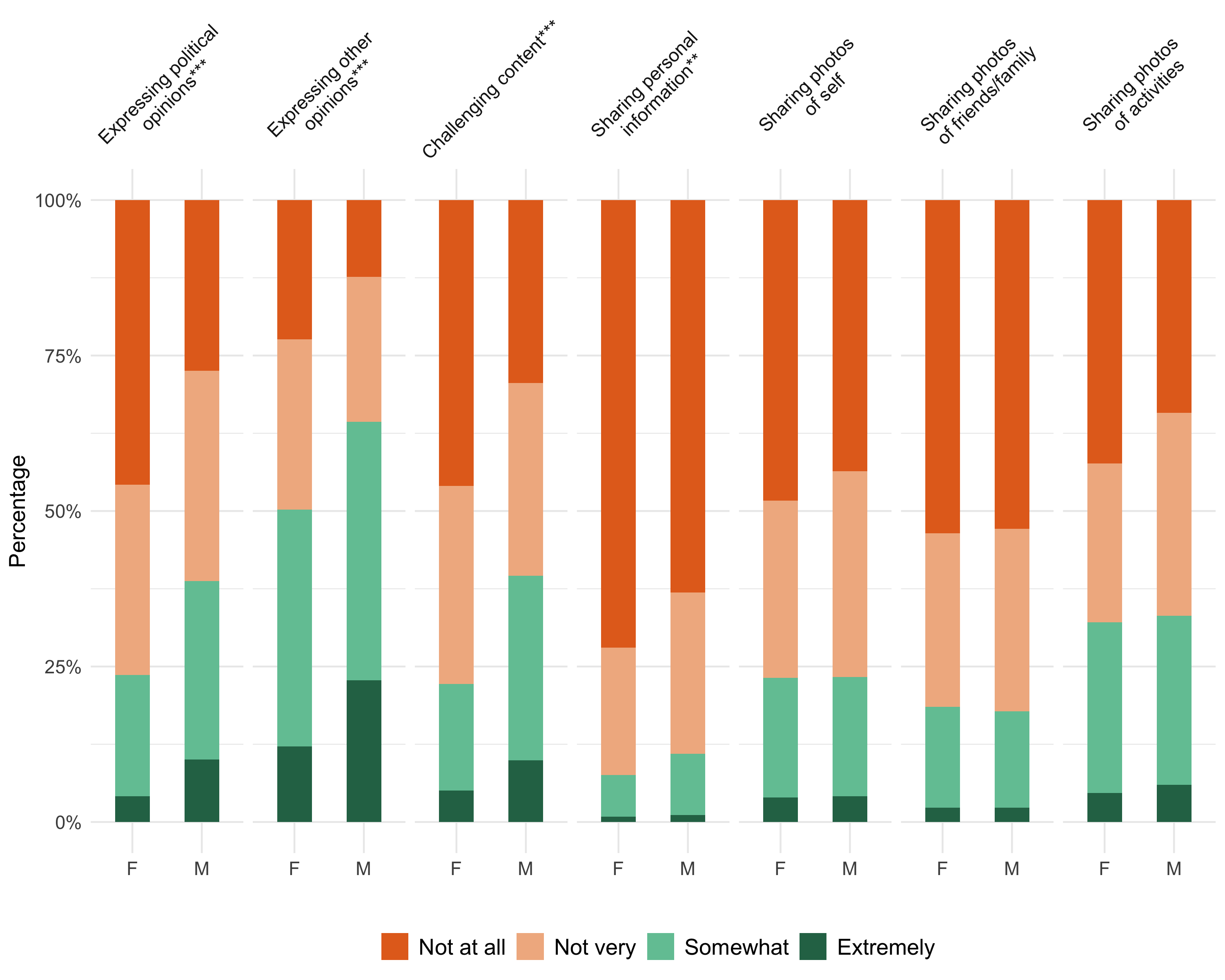}
    \caption{Comfort with seven online behaviours by gender differences. *** indicates a significant difference at p$<$.001; ** indicates a significant difference at p$<$.01.}
    \label{fig:comfort_level}
\end{figure*}

Just 23\% of women are comfortable expressing political opinions online compared to almost 40\% of men, 50\% of women are comfortable expressing other opinions online compared to 64\% of men, 22\% of women are comfortable challenging content they disagree with online compared to 40\% of men, and 7.5\% of women are comfortable sharing personal information online compared to 11\% of men. Logistic regressions showed women are 51\% less comfortable than men expressing political opinions online, 44\% less comfortable than men expressing other opinions online, and 57\% less comfortable than men challenging content they disagree with online (all ps$<$.001). Women are also 34\% less comfortable sharing personal information online (p=.008). There were no gender differences in comfort levels for sharing photos of self, of friends and family, and of activities (all ps$>$.05). Figure \ref{fig:comfort_level} shows self-reported comfort levels with the seven online behaviours. 

\subsection{Associations between fear of harm and comfort with online behaviours}
To test for relationships between fears about online harms and comfort engaging in several forms of online participation, we ran a series of 84 logistic regression analyses, each including fear as predictor (one for fear relating to each of 12 harms), and comfort engaging in the online behaviour of interest as the outcome (one for comfort with each of the seven behaviours), all controlling for gender, age, and overall internet use. Instead of including a regression model for all fifteen harms, we excluded fear of receiving misinformation, catfishing, and eating disorder content, where there is less theoretical basis to suppose an association with comfort with the seven behaviours of interest. The remaining 12 harms we included are mostly considered `contact' harms where there is risk of being directly targeted by harmful behaviours. Hate speech and misogyny may be considered `content' harms in general, but still pose risk as contact harms if respondents' social identities are targeted. In each model, we used a binary predictor for fear of each harm and a binary outcome for comfort with each behaviour, as described in the relevant sections above. 

\begin{figure*}[ht]
    \centering
    \includegraphics[width=1\linewidth]{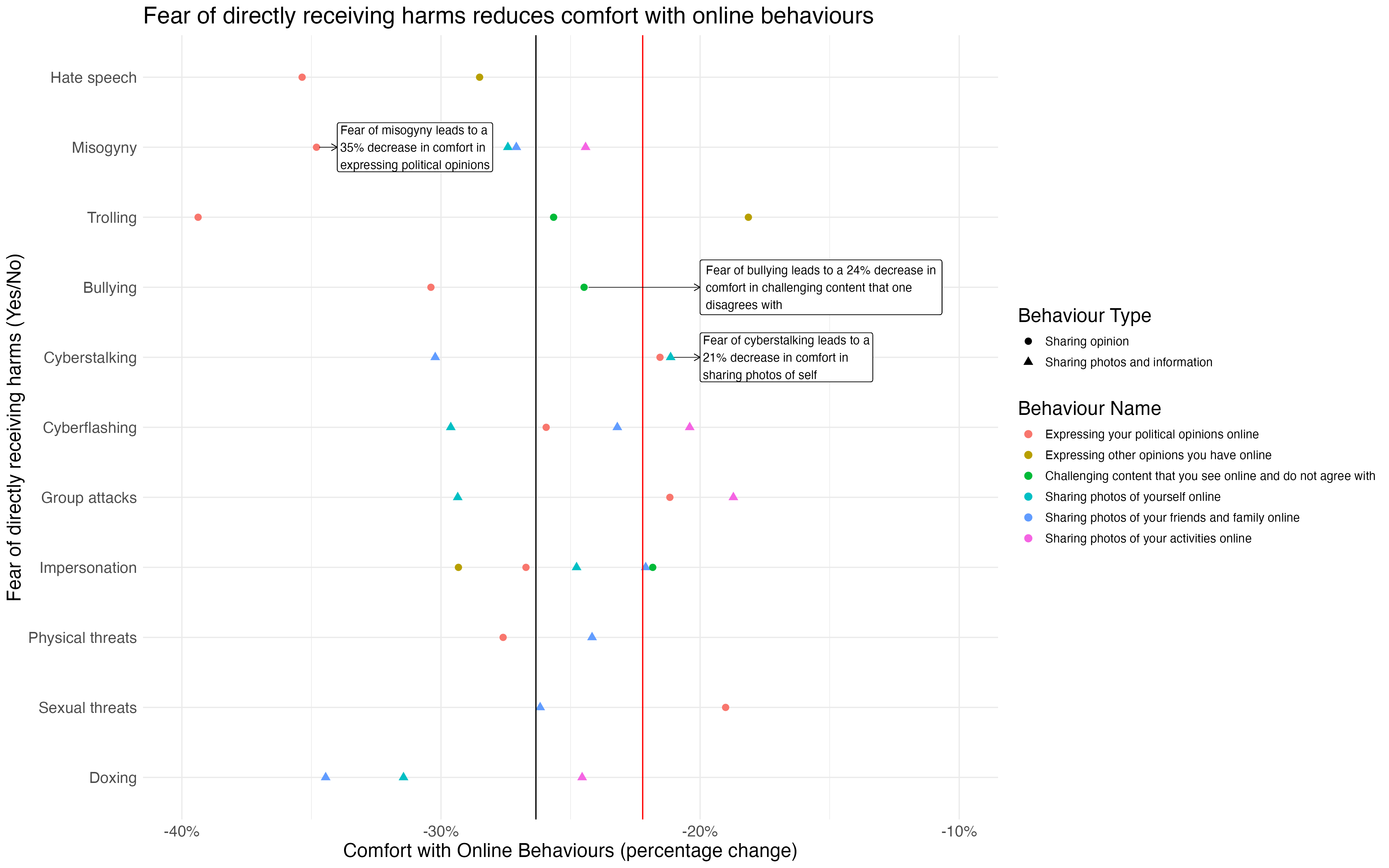}
    \caption{Fear of directly receiving harms is associated with reduced comfort with online behaviours. Here, significant associations between fears and comfort with behaviours are shown in different coloured (and shaped) points for each online behaviour. The black line highlights the average percentage change in comfort with online participation associated with fear of harm, and the red line highlights the average percentage change in comfort with online participation associated with gender. Sharing personal information is not included as none of these models were significant.}
    \label{fig:explanatory model}
\end{figure*}

Results from our regression analyses are shown in Figure \ref{fig:explanatory model}, with significant associations between fears and behaviours shown as individual points (non-significant associations not included). Each behaviour (shown in coloured points, with circles for the cluster of behaviours involving sharing opinions, and triangles for the cluster of behaviours involving sharing photos and personal information) is regressed onto each fear (y-axis), with the x-axis displaying the percentage decrease in comfort with each behaviour as a result of fear. 

On average, being fearful of receiving online harms is associated with being 26\% less comfortable with sharing opinions, photos or other information online. Furthermore, fear of different harms impacts on comfort participating with different online behaviours. For example, fear of being targeted by trolling is associated with reduced comfort with sharing political opinions, other opinions, and challenging content. Fear of being targeted by misogyny, cyberstalking and doxing is associated with reduced comfort with sharing photos. Full results for each regression model can be found in Table \ref{tab:Regressions} in Supplementary Information. 

\section{Discussion}
Using a nationally representative survey of UK adults, we examined the role of gender in shaping exposure to harmful online content, fears surrounding such exposure, the psychological impacts of online experiences, engagement with safety tools, and comfort with online participation. 

When asked about whether they had seen different forms of harmful online content over the last year, men and women reported similar levels of exposure overall, consistent with prior work asking broadly about how much people have seen general content which they consider to be harmful online \citep{Turing2023_harmstracker}. However, by asking about fifteen specific harms and the capacity in which people have seen them, we were able to detect more nuance in experiences. Here, women report being directly targeted by certain forms of `contact-based' harms like cyberflashing, cyberstalking and image based abuse to a greater extent than men, consistent with related prior work  \citep{Storry&Poppleton2022, UN2018, Amnesty2017}. These findings align with feminist accounts of the `continuum of violence' \citep{Kelly1988}, supporting the notion that digital abuse and harassment are part of the broader spectrum of gendered violence that extends across offline and online contexts. Our findings highlight that digital spaces are not neutral, and rather reflect and reproduce existing structural inequalities. 

However, we also found that men report being directly targeted by threats of physical violence, hate speech and trolling to a greater extent than women. These results are more novel and suggest that the types of harm men and women experience online reflect offline gendered norms, where women more commonly experience sexualised and silencing harms that reproduce structural subordination, while men may encounter more overt, confrontational harms. Our findings show that our approach in including men and women in our sample and asking about exposure to a range of harms is beneficial in gaining a more nuanced picture of gender differences in exposure.

We also found that women report having experienced greater negative psychological impact than men as a result of certain experiences online, including feeling sad and low, angry and frustrated, and with physical symptoms such as insomnia or headaches. Our results extend prior work showing that women may be more negatively affected by online harms than men \citep{Storry&Poppleton2022} by demonstrating increased negative affect in a representative sample of women, not only those who have been direct targets of abuse or harassment. These findings align with feminist accounts of `symbolic violence’ \cite{bourdieu1991language}, where routine exposure to gender-based inequality reinforces women's subordinate position and damages psychological wellbeing. 

Consistent with our finding that women report more negative psychological impact as a result of online experiences than men, we found that women expressed greater fear of being targeted by all fifteen harms that we asked about compared to men. These results show that women carry an additional psychological burden relating to safety concerns online, according with recent survey work showing that women are significantly more fearful than men of becoming the subjects of harmful deepfakes, such as deepfake pornography \citep{sippy2024behind}. Our findings also accord with work in the offline crime literature showing that women typically fear for their personal safety more than men do \citep{snedker2012explaining}. Feminist scholarship has long suggested that women’s disproportionate fear of violence reflects their structural positioning in a gendered world \citep{stanko1990everyday, Kelly1988}. Through this lens, the heightened fears reported by women reflect not only individual risk perception, but broader patterns of gendered safety work, where women carry the burden of responsibility for protecting themselves \citep{VeraGrayKelly2020}. Both men and women indicated that fear of exposure to online harms most typically originate from the media and from female friends’ experiences, suggesting that along with the role of the media in perpetuating safety concerns \citep[consistent with][]{cashmore2014fear}, the known negative experiences of other women are also particularly salient.

It is interesting to consider the interplay between fear and exposure. While women consistently express greater fear about being targeted by harms, they are not always at greater risk of such exposure. Feminist criminology highlights that women's fears surrounding personal harm often exceed their actual likelihood of experiencing such harm, with fear a mechanism in the structural oppression of women, functioning to regulate women's behaviour and restrict participation \citep{stanko1990everyday, Kelly1988}. It is important to question whether increased fears lead women to do more to protect themselves online in the form of safety work. At the same time, we cannot rule out that women are simply more comfortable expressing fear than men. 

In line with our conjecture that women may do more safety work to protect themselves from harmful online experiences than men, we found that women report using all seven safety tools that we asked about to a greater extent than men. These results align with theoretical claims that women engage in disproportionate self-protection work \citep{VeraGrayKelly2020}. Our findings suggest that the unequal patterns of safety work long observed offline are reproduced in online spaces, reinforcing the additional burden of risk management for women. 

As well as engaging with safety tools to a great extent than men, women also report being less comfortable than men with participating in several online behaviours including expressing political opinions online, expressing opinions more generally, and challenging content that they disagree with. While some prior work shows that women conduct themselves differently online after experiencing abuse \citep{Amnesty2017}, our findings are novel in directly comparing men and women's general comfort with online participation. Our results suggest that women may be self-censoring online, particularly in terms of expressing opinions and challenging other content. This can be understood as a form of gendered silencing, where fears around harm exposure interact with disproportionate safety work to constrain women's digital citizenship and public voice \citep{vera2018right, olson2018combating, chen2020you}. These results are concerning, because with much public discourse happening online, gender inequality in public spaces is likely to be perpetuated if women feel too fearful to participate. 

Our findings suggest that greater fear surrounding exposure to online harms plays a key role in women's self-censorship online. Women were more likely than men to say that they are less likely to share opinions, photos and general content as a direct result of fearing being targeted by online harms. These patterns accord with offline behaviours whereby fear of crime leads to moderation of one's actions, such as avoiding walking in quiet places after dark, or taking extra precautions when travelling alone \citep{ONS2022}. Consistent with this, our regression analysis showed that greater fear of harm exposure was associated with lower comfort with several different online behaviours, even when controlling for gender, age, and internet use. For example, fear of directly receiving online misogyny significantly predicted less comfort expressing political opinions online, while fear of cyberstalking significantly predicted less comfort sharing photos of the self and of friends and family online. The interplay between harm exposure, fear, safety work, and participation is likely to be complex. It may be that greater exposure to particular harms increases fear in women, leading to additional safety work, or that greater fears arise by factors such as cultural transmission through media and social networks. Future research would benefit from working to understand the causal nature of these relationships in more detail, which could inform how to best target intervention. 

While our work offers important novel insights into the differential impacts of online harms experienced by men and women, it is important to acknowledge limitations in the study and outstanding research questions that were not possible to answer with this data. Our Prolific sample was designed to be representative of the UK population across age, gender, and ethnicity, but because Prolific recruitment is not probability-based and relies on participants filling survey slots on a first-come, first-served basis, there remains the possibility of selection bias.

Additionally, in attempting to understand gendered experiences of online harms, we statistically compare responses given by people identifying as women with those identifying as men. In our sample, a small number of respondents identified as non-binary or another gender (N=17 total), but the sample size was too limited to allow for robust statistical comparisons. Future work could look more specifically at the experiences of individuals identifying in other ways by oversampling this group and actively engaging this sample to capture distinct experiences of online harms. Additionally, our sample included only adults, and future work would benefit from tackling similar research questions with children and teenagers \citep{Plan2020, livingstone2014risk}.   

In our analyses, we generalise across men and women as a whole, but there are likely to be intersections with gender and other characteristic such as age and ethnicity. Future work should consider these intersections in greater detail, for example in examining differing experiences of older vs. younger women and men, or exploring gender differences based on ethnicity, for example extending findings from \cite{OfCom2022} which suggested Black women were least likely to think that being online has a positive effect on mental health than those from other ethnic backgrounds. 

Overall, we provide novel evidence about how and when women are disproportionately affected by online harms. Our findings demonstrate the myriad of harms that men and women are exposed to online, providing valuable information about who is most at risk and when, important for knowing where psychological support should most be directed and in what ways interventions against online harms may be most effective. Our work highlights the need for greater efforts from platforms to protect women users, welcoming efforts from the Online Safety Act in requiring Ofcom to develop guidance for tech companies to reduce harm to women and girls \citep{EVAW2023, Ofcom2025SaferLife}, but we also emphasise that such measures must go beyond individual focus and address the systemic reproduction of inequality in digital spaces. To ensure an egalitarian society, we must make sure that all members of society feel safe and able to participate in online spaces.

\section*{Corresponding author:}
\label{Author contact}
Florence E. Enock, fenock@turing.ac.uk 

%\section*{Open science}
%\label{Open science}

\section*{Acknowledgements}
This work was supported by the Ecosystem Leadership Award under the EPSRC Grant EPX03870X1 and The Alan Turing Institute. 

%% The Appendices part is started with the command \appendix;
%% appendix sections are then done as normal sections
%%\appendix

%%\section{Appendix title 1}
%% \label{}

%%\section{Appendix title 2}
%% \label{}

%% If you have bibdatabase file and want bibtex to generate the
%% bibitems, please use
%%
\bibliographystyle{apalike} 
\bibliography{Survey3_citations}

%% else use the following coding to input the bibitems directly in the
%% TeX file.

%%\begin{thebibliography}{00}

%% \bibitem[Author(year)]{label}
%% For example:

%% \bibitem[Aladro et al.(2015)]{Aladro15} Aladro, R., Martín, S., Riquelme, D., et al. 2015, \aas, 579, A101

%%\end{thebibliography}

\section{Supplementary Information}
\label{SI}

\subsection{Supplementary tables}

\subsubsection{Definitions of online harms}
\begin{table}
\begin{tabular}{>{\raggedright\arraybackslash}p{0.25\linewidth}>{\raggedright\arraybackslash}p{0.75\linewidth}} 
\textit{\textbf{Harm}} & \textit{\textbf{Definition}}\\ 
Misinformation& Misinformation is false or inaccurate information, regardless of whether there is intent to mislead.\\ 
 Online misogyny&  Online misogyny is content expressing hatred for, contempt of, or prejudice against women and girls.\\ 
 Online trolling&  Online trolling is the practice of leaving intentionally provocative or offensive messages on the internet in order to get attention, cause trouble, upset or provoke reactions from other online users.\\ 
 Online bullying&  Online bullying is online intimidation which may include repeatedly sending cruel text or images to a victim aimed to scare, anger or shame them.\\ 
 Cyberstalking& Cyberstalking is persistent and unwanted attention online that makes the victim feel pestered and harassed, such as repeated messages on social media or dating apps.\\ 
 Cyberflashing&  Cyberflashing is sending unwanted graphic nude images to people without their consent.\\ 
 Group attacks&  A group attack is when a group of users shame, boycott or exclude another user because of their views, including ‘pile-ons’, when a group of people gang up on a single victim.\\ 
 Online impersonation&  Online impersonation is the practice of creating a fake profile pretending to be the victim.\\ 
 Catfishing&  Catfishing is deception in which a person creates a fictional persona or fake identity online, usually in order to compromise a specific victim in some way.\\ 
 Online threats of non-sexual physical violence&  Online threats of non-sexual physical violence is the sending or posting of threats of violence to an individual or group of individuals, including death threats, threats to stab, shoot and harm, and threats to harm family members.\\ 
 Online threats of sexual violence& Online threats of sexual violence is the sending or posting of threats of sexual abuse to an individual or group of individuals, including rape threats and similar threats to family members.\\ 
 Doxing&  Doxing is the action or process of publishing private or identifying information about a particular individual on the internet, typically with malicious intent.\\ 
 Image based sexual abuse&  Image based sexual abuse is the sharing of intimate images taken with or without someone’s consent and shared without their consent or knowledge (often referred to as 'revenge porn’).\\ 
 Eating disorder content&  Eating disorder content is content which glorifies unhealthy body image or eating disorders, often promoting excessive dieting and exercise.\\ \end{tabular}
    \label{tab:definitions}
\end{table}

\subsubsection{Significant gender differences in directly receiving online harms}
\begin{table}
\begin{tabular}{llll}
\textit{\textbf{Harm}} & \textit{\textbf{Gender}} & \textit{\textbf{Response}} & \textit{\textbf{\%}} \\ \hline
Misogyny & Female & Never & 87.0\% \\
 &  & Once & 5.9\% \\
 &  & 2-4 times & 4.7\% \\
 &  & 5+ times & 2.5\% \\
 & Male & Never & 95.0\% \\
 &  & Once & 1.9\% \\
 &  & 2-4 times & 1.8\% \\
 &  & 5+ times & 1.3\% \\ \hline
Cyberstalking & Female & Never & 96.6\% \\
 &  & Once & 2.3\% \\
 &  & 2-4 times & 0.6\% \\
 &  & 5+ times & 0.5\% \\
 & Male & Never & 98.2\% \\
 &  & Once & 0.9\% \\
 &  & 2-4 times & 0.6\% \\
 &  & 5+ times & 0.2\% \\ \hline
Cyberflashing & Female & Never & 95.0\% \\
 &  & Once & 2.2\% \\
 &  & 2-4 times & 1.8\% \\
 &  & 5+ times & 1.1\% \\
 & Male & Never & 97.8\% \\
 &  & Once & 0.9\% \\
 &  & 2-4 times & 0.6\% \\
 &  & 5+ times & 0.6\% \\ \hline
\multirow{2}{*}{\begin{tabular}[c]{@{}l@{}}Eating disorder \\ content\end{tabular}} & Female & Never & 95.8\% \\
 &  & Once & 1.6\% \\
 &  & 2-4 times & 1.9\% \\
 &  & 5+ times & 0.7\% \\
 & Male & Never & 97.5\% \\
 &  & Once & 1.1\% \\
 &  & 2-4 times & 0.8\% \\
 &  & 5+ times & 0.5\% \\ \hline
\multirow{2}{*}{\begin{tabular}[c]{@{}l@{}}Image-based \\ abuse\end{tabular}} & Female & Never & 98.4\% \\
 &  & Once & 1.0\% \\
 &  & 2-4 times & 0.5\% \\
 &  & 5+ times & 0.1\% \\
 & Male & Never & 99.3\% \\
 &  & Once & 0.5\% \\
 &  & 2-4 times & 0.2\% \\
 &  & 5+ times & 0.0\%
\end{tabular}
 \caption{The extent to which men and women say they have been directly targeted by each of the five online harms that women report being targeted by to a significantly greater extent than men. Percentages are representative of the population. `Prefer not to say' responses were uncommon and are not included.}
    \label{tab:Exposure_DR_women}
\end{table}

\begin{table}
\begin{tabular}{llll}
\textbf{Harm} & \textbf{Gender} & \textbf{Response} & \textbf{\%} \\ \hline
Hate speech & Female & Never & 89.1\% \\
 &  & Once & 4.5\% \\
 &  & 2-4 times & 4.9\% \\
 &  & 5+ times & 1.6\% \\
 & Male & Never & 84.2\% \\
 &  & Once & 5.5\% \\
 &  & 2-4 times & 6.8\% \\
 &  & 5+ times & 3.5\% \\ \hline
Misinformation & Female & Never & 71.4\% \\
 &  & Once & 9.6\% \\
 &  & 2-4 times & 11.9\% \\
 &  & 5+ times & 7.0\% \\
 & Male & Never & 60.4\% \\
 &  & Once & 8.6\% \\
 &  & 2-4 times & 15.3\% \\
 &  & 5+ times & 15.7\% \\ \hline
Trolling & Female & Never & 89.1\% \\
 &  & Once & 4.6\% \\
 &  & 2-4 times & 3.8\% \\
 &  & 5+ times & 2.6\% \\
 & Male & Never & 77.6\% \\
 &  & Once & 5.6\% \\
 &  & 2-4 times & 9.2\% \\
 &  & 5+ times & 7.6\% \\ \hline
\multirow{2}{*}{\begin{tabular}[c]{@{}l@{}}Threats of \\ physical violence\end{tabular}} & Female & Never & 96.5\% \\
 &  & Once & 1.7\% \\
 &  & 2-4 times & 1.1\% \\
 &  & 5+ times & 0.7\% \\
 & Male & Never & 94.0\% \\
 &  & Once & 2.7\% \\
 &  & 2-4 times & 2.1\% \\
 &  & 5+ times & 1.2\%
\end{tabular}
 \caption{The extent to which men and women say they have been directly targeted by each of the four online harms that men report being targeted by to a significantly greater extent than women. Percentages are representative of the population. `Prefer not to say' responses were uncommon and are not included.}
    \label{tab:Exposure_DR_men}
\end{table}

\subsubsection{Gender differences in psychological impacts of online experiences}

\begin{table}
\begin{tabular}{llll}
Feeling                & Gender & Response      & \%     \\ \hline
Sad or low             & Female & Not at all    & 15.8\% \\
                       &        & Not very much & 28.1\% \\
                       &        & Somewhat      & 45.8\% \\
                       &        & Very much     & 10.2\% \\
                       & Male   & Not at all    & 24.3\% \\
                       &        & Not very much & 36.3\% \\
                       &        & Somewhat      & 32.7\% \\
                       &        & Very much     & 6.7\%  \\ \hline
Angry or frustrated    & Female & Not at all    & 9.6\%  \\
                       &        & Not very much & 17.7\% \\
                       &        & Somewhat      & 47.8\% \\
                       &        & Very much     & 24.9\% \\
                       & Male   & Not at all    & 13.8\% \\
                       &        & Not very much & 22.8\% \\
                       &        & Somewhat      & 45.7\% \\
                       &        & Very much     & 17.6\% \\ \hline
Physically affected    & Female & Not at all    & 62.2\% \\
                       &        & Not very much & 24.8\% \\
                       &        & Somewhat      & 10.6\% \\
                       &        & Very much     & 2.4\%  \\
                       & Male   & Not at all    & 71.8\% \\
                       &        & Not very much & 19.0\% \\
                       &        & Somewhat      & 8.2\%  \\
                       &        & Very much     & 1.0\%  \\ \hline
Job affected           & Female & Not at all    & 70.3\% \\
                       &        & Not very much & 19.7\% \\
                       &        & Somewhat      & 8.2\%  \\
                       &        & Very much     & 1.8\%  \\
                       & Male   & Not at all    & 72.0\% \\
                       &        & Not very much & 20.0\% \\
                       &        & Somewhat      & 6.1\%  \\
                       &        & Very much     & 1.9\%  \\ \hline
Relationships affected & Female & Not at all    & 61.0\% \\
                       &        & Not very much & 23.5\% \\
                       &        & Somewhat      & 12.8\% \\
                       &        & Very much     & 2.8\%  \\
                       & Male   & Not at all    & 62.2\% \\
                       &        & Not very much & 23.5\% \\
                       &        & Somewhat      & 11.4\% \\
                       &        & Very much     & 2.8\%  \\ \hline
Motivated to act       & Female & Not at all    & 47.6\% \\
                       &        & Not very much & 28.8\% \\
                       &        & Somewhat      & 20.4\% \\
                       &        & Very much     & 3.2\%  \\
                       & Male   & Not at all    & 49.3\% \\
                       &        & Not very much & 30.3\% \\
                       &        & Somewhat      & 17.2\% \\
                       &        & Very much     & 3.2\% 
\end{tabular}
 \caption{The extent to which people say they have experienced each kind of psychological impact as a result of online experiences split by gender.}
    \label{tab:Feelings}
\end{table}

\subsubsection{Associations between fear of each online harm and comfort with each online activity}

\onecolumn
\begin{landscape}
\begin{longtable}[c]{lllllllll}
\caption{Coefficients for each regression model examining associations between fear of each online harm and comfort with each online activity, controlling for age, gender and internet use frequency. We show estimates, exponenciated coefficients and p-values before and before and after influential values are identified and removed. Note that influential values are identified using Cook's distance, with a threshold of 4/N. When p=0.000, this means p$<$.001.}
\label{tab:Regressions}\\
\textbf{Fear (IV)} &
  \textbf{Behaviour (DV)} &
  \textbf{Estimate} &
  \textbf{Exp. coefficient} &
  \textit{\textbf{p}} &
  \textbf{\begin{tabular}[c]{@{}l@{}}N inf. \\ values\end{tabular}} &
  \textbf{\begin{tabular}[c]{@{}l@{}}Estimate,\\ inf. values \\ removed\end{tabular}} &
  \textbf{\begin{tabular}[c]{@{}l@{}}Exp. coefficient,\\ inf. values \\ removed\end{tabular}} &
  \textit{\textbf{\begin{tabular}[c]{@{}l@{}}p,\\ inf. values \\ removed\end{tabular}}} \\
\endfirsthead
\multicolumn{9}{c}%
{{\bfseries Table \thetable\ continued from previous page}} \\
\textbf{Fear (IV)} &
  \textbf{Behaviour (DV)} &
  \textbf{Estimate} &
  \textbf{Exp. coefficient} &
  \textit{\textbf{p}} &
  \textbf{\begin{tabular}[c]{@{}l@{}}N inf. \\ values\end{tabular}} &
  \textbf{\begin{tabular}[c]{@{}l@{}}Estimate,\\ inf. values \\ removed\end{tabular}} &
  \textbf{\begin{tabular}[c]{@{}l@{}}Exp. coefficient,\\ inf. values \\ removed\end{tabular}} &
  \textit{\textbf{\begin{tabular}[c]{@{}l@{}}p,\\ inf. values \\ removed\end{tabular}}} \\
\endhead
Hate speech      & Expressing political opinions        & -0.44 & 0.65 & 0.000 & 39  & -0.64  & 0.53 & 0.000 \\
                 & Expressing other opinions            & -0.34 & 0.72 & 0.001 & 40  & -0.35  & 0.70 & 0.000 \\
                 & Sharing personal information         & 0.00  & 1.00 & 0.982 & 169 & -18.35 & 0.00 & 0.995 \\
                 & Sharing photos of self               & -0.13 & 0.88 & 0.295 & 89  & -0.26  & 0.77 & 0.055 \\
                 & Sharing photos of friends and family & -0.09 & 0.91 & 0.486 & 139 & -0.10  & 0.91 & 0.556 \\
                 & Sharing photos of activities         & -0.13 & 0.88 & 0.224 & 50  & -0.21  & 0.81 & 0.065 \\
                 & Challenging content                  & -0.18 & 0.83 & 0.096 & 45  & -0.28  & 0.75 & 0.014 \\
Misogyny         & Expressing political opinions        & -0.43 & 0.65 & 0.000 & 40  & -0.67  & 0.51 & 0.000 \\
                 & Expressing other opinions            & -0.18 & 0.84 & 0.093 & 39  & -0.19  & 0.83 & 0.074 \\
                 & Sharing personal information         & 0.07  & 1.08 & 0.692 & 159 & -17.06 & 0.00 & 0.992 \\
                 & Sharing photos of self               & -0.32 & 0.73 & 0.015 & 102 & -0.68  & 0.51 & 0.000 \\
                 & Sharing photos of friends and family & -0.32 & 0.73 & 0.027 & 150 & -0.82  & 0.44 & 0.000 \\
                 & Sharing photos of activities         & -0.28 & 0.76 & 0.016 & 56  & -0.43  & 0.65 & 0.001 \\
                 & Challenging content                  & -0.20 & 0.82 & 0.099 & 46  & -0.31  & 0.74 & 0.016 \\
Trolling         & Expressing political opinions        & -0.50 & 0.61 & 0.000 & 31  & -0.64  & 0.53 & 0.000 \\
                 & Expressing other opinions            & -0.20 & 0.82 & 0.038 & 40  & -0.23  & 0.80 & 0.020 \\
                 & Sharing personal information         & 0.20  & 1.22 & 0.230 & 173 & -17.91 & 0.00 & 0.994 \\
                 & Sharing photos of self               & -0.16 & 0.85 & 0.162 & 97  & -0.23  & 0.80 & 0.095 \\
                 & Sharing photos of friends and family & -0.06 & 0.94 & 0.614 & 136 & -0.05  & 0.95 & 0.760 \\
                 & Sharing photos of activities         & -0.08 & 0.92 & 0.418 & 46  & -0.13  & 0.88 & 0.232 \\
                 & Challenging content                  & -0.30 & 0.74 & 0.006 & 44  & -0.37  & 0.69 & 0.001 \\
Bullying         & Expressing political opinions        & -0.36 & 0.70 & 0.001 & 32  & -0.48  & 0.62 & 0.000 \\
                 & Expressing other opinions            & -0.12 & 0.89 & 0.221 & 39  & -0.14  & 0.87 & 0.148 \\
                 & Sharing personal information         & 0.04  & 1.04 & 0.818 & 177 & -18.54 & 0.00 & 0.996 \\
                 & Sharing photos of self               & -0.11 & 0.90 & 0.338 & 89  & -0.23  & 0.79 & 0.077 \\
                 & Sharing photos of friends and family & -0.14 & 0.87 & 0.268 & 135 & -0.12  & 0.89 & 0.438 \\
                 & Sharing photos of activities         & -0.01 & 0.99 & 0.930 & 44  & -0.06  & 0.94 & 0.572 \\
                 & Challenging content                  & -0.28 & 0.76 & 0.008 & 43  & -0.36  & 0.70 & 0.001 \\
Cyberstalking    & Expressing political opinions        & -0.24 & 0.79 & 0.020 & 25  & -0.31  & 0.74 & 0.004 \\
                 & Expressing other opinions            & -0.01 & 0.99 & 0.938 & 40  & -0.04  & 0.97 & 0.724 \\
                 & Sharing personal information         & -0.19 & 0.83 & 0.258 & 176 & -18.93 & 0.00 & 0.996 \\
                 & Sharing photos of self               & -0.24 & 0.79 & 0.041 & 95  & -0.43  & 0.65 & 0.001 \\
                 & Sharing photos of friends and family & -0.36 & 0.70 & 0.004 & 136 & -0.34  & 0.71 & 0.034 \\
                 & Sharing photos of activities         & -0.12 & 0.89 & 0.238 & 46  & -0.17  & 0.85 & 0.115 \\
                 & Challenging content                  & -0.16 & 0.85 & 0.123 & 39  & -0.22  & 0.81 & 0.045 \\
Cyberflashing    & Expressing political opinions        & -0.30 & 0.74 & 0.008 & 30  & -0.43  & 0.65 & 0.000 \\
                 & Expressing other opinions            & -0.12 & 0.89 & 0.261 & 38  & -0.13  & 0.88 & 0.228 \\
                 & Sharing personal information         & 0.06  & 1.07 & 0.713 & 167 & -18.37 & 0.00 & 0.995 \\
                 & Sharing photos of self               & -0.35 & 0.70 & 0.005 & 97  & -0.59  & 0.56 & 0.000 \\
                 & Sharing photos of friends and family & -0.26 & 0.77 & 0.048 & 141 & -0.30  & 0.74 & 0.079 \\
                 & Sharing photos of activities         & -0.23 & 0.80 & 0.038 & 49  & -0.33  & 0.72 & 0.005 \\
                 & Challenging content                  & -0.07 & 0.93 & 0.512 & 40  & -0.14  & 0.87 & 0.241 \\
Group attacks    & Expressing political opinions        & -0.24 & 0.79 & 0.024 & 30  & -0.34  & 0.71 & 0.002 \\
                 & Expressing other opinions            & -0.09 & 0.92 & 0.379 & 39  & -0.10  & 0.91 & 0.327 \\
                 & Sharing personal information         & -0.01 & 0.99 & 0.947 & 177 & -18.67 & 0.00 & 0.997 \\
                 & Sharing photos of self               & -0.35 & 0.71 & 0.003 & 99  & -0.58  & 0.56 & 0.000 \\
                 & Sharing photos of friends and family & -0.22 & 0.80 & 0.085 & 138 & -0.26  & 0.77 & 0.108 \\
                 & Sharing photos of activities         & -0.21 & 0.81 & 0.047 & 48  & -0.27  & 0.76 & 0.013 \\
                 & Challenging content                  & -0.13 & 0.88 & 0.208 & 42  & -0.20  & 0.82 & 0.066 \\
Impersonation    & Expressing political opinions        & -0.31 & 0.73 & 0.002 & 21  & -0.34  & 0.71 & 0.001 \\
                 & Expressing other opinions            & -0.35 & 0.71 & 0.000 & 39  & -0.35  & 0.70 & 0.000 \\
                 & Sharing personal information         & -0.28 & 0.75 & 0.078 & 178 & -18.68 & 0.00 & 0.996 \\
                 & Sharing photos of self               & -0.29 & 0.75 & 0.012 & 96  & -0.44  & 0.64 & 0.001 \\
                 & Sharing photos of friends and family & -0.25 & 0.78 & 0.040 & 135 & -0.29  & 0.75 & 0.059 \\
                 & Sharing photos of activities         & -0.08 & 0.93 & 0.432 & 45  & -0.08  & 0.92 & 0.435 \\
                 & Challenging content                  & -0.25 & 0.78 & 0.015 & 39  & -0.28  & 0.76 & 0.007 \\
Physical threats & Expressing political opinions        & -0.32 & 0.72 & 0.002 & 27  & -0.39  & 0.67 & 0.000 \\
                 & Expressing other opinions            & -0.14 & 0.87 & 0.142 & 38  & -0.15  & 0.86 & 0.126 \\
                 & Sharing personal information         & -0.04 & 0.96 & 0.788 & 175 & -17.91 & 0.00 & 0.994 \\
                 & Sharing photos of self               & -0.20 & 0.82 & 0.089 & 99  & -0.36  & 0.70 & 0.007 \\
                 & Sharing photos of friends and family & -0.28 & 0.76 & 0.028 & 136 & -0.25  & 0.78 & 0.125 \\
                 & Sharing photos of activities         & -0.09 & 0.92 & 0.385 & 43  & -0.09  & 0.92 & 0.404 \\
                 & Challenging content                  & -0.12 & 0.88 & 0.237 & 39  & -0.17  & 0.85 & 0.125 \\
Sexual threats   & Expressing political opinions        & -0.21 & 0.81 & 0.050 & 26  & -0.29  & 0.75 & 0.010 \\
                 & Expressing other opinions            & -0.11 & 0.90 & 0.282 & 38  & -0.12  & 0.89 & 0.234 \\
                 & Sharing personal information         & -0.17 & 0.84 & 0.318 & 168 & -18.41 & 0.00 & 0.994 \\
                 & Sharing photos of self               & -0.20 & 0.82 & 0.089 & 97  & -0.40  & 0.67 & 0.004 \\
                 & Sharing photos of friends and family & -0.30 & 0.74 & 0.019 & 144 & -0.49  & 0.61 & 0.004 \\
                 & Sharing photos of activities         & -0.03 & 0.97 & 0.778 & 44  & -0.05  & 0.95 & 0.654 \\
                 & Challenging content                  & -0.17 & 0.85 & 0.126 & 41  & -0.21  & 0.81 & 0.061 \\
Doxing           & Expressing political opinions        & -0.17 & 0.84 & 0.093 & 25  & -0.24  & 0.79 & 0.023 \\
                 & Expressing other opinions            & -0.03 & 0.97 & 0.785 & 39  & -0.04  & 0.96 & 0.706 \\
                 & Sharing personal information         & -0.18 & 0.83 & 0.260 & 182 & 0.00   & 1.00 & 1.000 \\
                 & Sharing photos of self               & -0.38 & 0.69 & 0.001 & 99  & -0.71  & 0.49 & 0.000 \\
                 & Sharing photos of friends and family & -0.42 & 0.66 & 0.001 & 137 & -0.62  & 0.54 & 0.000 \\
                 & Sharing photos of activities         & -0.28 & 0.75 & 0.006 & 50  & -0.37  & 0.69 & 0.001 \\
                 & Challenging content                  & -0.09 & 0.92 & 0.403 & 37  & -0.13  & 0.88 & 0.224 \\
\multirow{2}{*}{\begin{tabular}[c]{@{}l@{}}Image-based \\ sexual abuse\end{tabular}} &
  Expressing political opinions &
  -0.16 &
  0.86 &
  0.138 &
  22 &
  -0.20 &
  0.82 &
  0.066 \\
                 & Expressing other opinions            & -0.04 & 0.96 & 0.695 & 39  & -0.04  & 0.96 & 0.666 \\
                 & Sharing personal information         & 0.10  & 1.10 & 0.549 & 177 & -18.54 & 0.00 & 0.996 \\
                 & Sharing photos of self               & -0.10 & 0.91 & 0.402 & 94  & -0.25  & 0.78 & 0.063 \\
                 & Sharing photos of friends and family & -0.05 & 0.95 & 0.692 & 136 & -0.11  & 0.90 & 0.491 \\
                 & Sharing photos of activities         & 0.01  & 1.01 & 0.909 & 42  & 0.03   & 1.03 & 0.818 \\
                 & Challenging content                  & -0.09 & 0.92 & 0.405 & 40  & -0.11  & 0.90 & 0.331
\end{longtable}
\end{landscape}

\end{document}